\documentclass[paper,12pt]{JHEP3} 

\usepackage{graphicx}

\title{Deep inelastic scattering of baryons in a modified soft wall model}

\author{ Nelson R. F. Braga \\
Instituto de F\'{\i}sica,
Universidade Federal do Rio de Janeiro, Caixa Postal 68528, RJ
21941-972 -- Brazil\\}
\author{ Alfredo Vega \\ Departamento de F\'isica y Centro Cient\'ifico Tecnol\'ogico de Valpara\'iso (CCTVal), Universidad T\'ecnica Federico Santa Mar\'ia, Casilla 110-V, Valpara\'iso, Chile \\

E-mails: \email{braga@if.ufrj.br, alfredo.vega@usm.cl}}

\abstract{ We calculate the structure functions for unpolarized deep inelastic scattering of baryons using
an AdS/QCD soft wall model that considers a dressed mass for the bulk fermionic fields. 
Considering the regime of large Bjorken parameter x, we compare the results for the proton structure function 
$F_2 \,\,$ with experimental results.    }

 
\begin{document}
\maketitle

\section{ Introduction }

Our present knowledge of strong interactions indicates that QCD (Quantum Chromodynamics) is the quantum field theory that models the physics of this fundamental interaction. However, many important properties of the strongly interacting particles, the hadrons, can not be described using only perturbative QCD. This happens because the QCD coupling is large at low energies. Thus, in order to calculate static properties of hadrons, or to describe 
their structure,    one needs additional tools.

In the last years, many interesting models to study non perturbative aspects of hadronic physics were developed
based on the idea of gauge/string duality. The main source of inspiration for these kind of models was the discovery of the AdS/CFT correspondence\cite{Maldacena:1997re,Gubser:1998bc,Witten:1998qj} that is an exact duality between string theory in certain ten dimensional geometries and supersymmetric $ SU(N)$ gauge theories with large $N $ on the corresponding boundary. In particular, string theory in $AdS_5 \times S^5 $ space is dual to a four dimensional gauge theory.  
 
In the AdS/CFT correspondence, the gauge theory is conformal. The idea of breaking this conformal invariance by an Infrared (IR) cut off, represented in the dual AdS geometry by a geometrical cut off in the radial coordinate of the space, was introduced in \cite{Polchinski:2001tt} as a tool to reproduce the high energy scaling of hadronic scattering amplitudes for processes at fixed angles. This scaling had been found in QCD a long time before \cite{Matveev:1973ra,Brodsky:1973kr} but the corresponding string theory description was lacking for a very long time.   

This approach of considering a maximum radial size of AdS space as the dual of an infrared cut off 
in the gauge theory was then used in \cite{BoschiFilho:2002ta,BoschiFilho:2002vd} to calculate the mass spectrum of glueballs. In these articles, boundary conditions were imposed on fields living in an AdS slice and the
corresponding normal modes were associated with hadronic states. This kind of model, later called AdS/QCD hard wall, was then applied to other hadrons, as, for example, in \cite{deTeramond:2005su}.
It is important to mention the important  earlier works
 \cite{Csaki:1998qr,Hashimoto:1998if,Csaki:1998cb,Minahan:1998tm,Brower:2000rp}.
where glueball masses were calculated by considering an AdS Schwarzschild black hole as dual to a non-supersymmetric Yang Mills theory. 

The hadronic masses calculated using the hard wall model, for particles with a given spin, do not exhibit a  linear relation between mass squared and excitation number.
This motivated a different AdS/QCD approach consisting of a background involving AdS space plus a field that acts
effectively as a smooth  infrared cut off \cite{Karch:2006pv}.  This, so called soft wall model, leads to a linear relation between the mass squared and the radial excitation number for vector mesons (see also \cite{Colangelo:2007pt} for scalar glueballs). 
 
However, the soft wall model, as originally formulated\cite{Karch:2006pv}, does not work for fermions. That means, it does not lead to a discrete mass spectrum for fermions because the dilaton introduced in the action is factorized out in their equations of motion. In other words the fermions do not feel the smooth cut off of the soft wall model. Some alternative versions of the Soft Wall AdS/QCD models for fermions were developed then, with the purpose of reproducing the mass spectrum of baryons observed. 
One example can be found in \cite{Forkel:2007cm,dePaula:2008fp}, where the authors consider models in asymptotically AdS space including a warp factor in the metric. Other possibility, studied in 
\cite{Vega:2008te,Abidin:2009hr,Gutsche:2011vb}, considers a z dependent (or dressed)  mass for the fermionic modes propagating in AdS space. This last approach has been considered in \cite{Abidin:2009hr} to study nucleon form factors, and in \cite{Vega:2010ns} to obtain some generalized parton distributions (GPD) for nucleons. 

It is important to note that experimental data for the mass spectrum of baryons of spin 1/2 show that the square of the masses of the excited states are almost equally spaced. This can be seen, for example, in  \cite{Klempt:2002vp}  (see specially table II in this reference). That means, there is an approximately linear relation between the mass squared and the excitation level quantum number for these baryons of spin 1/2.

The study of deep inelastic scattering (DIS) using the AdS/CFT correspondence appeared first in \cite{Polchinski:2002jw}. Then other authors considered the description of DIS using various AdS/QCD models like, for example,  in \cite{BallonBayona:2007qr,BallonBayona:2007rs,Cornalba:2008sp,Pire:2008zf,Albacete:2008ze,Gao:2009ze,
Hatta:2009ra,BallonBayona:2008zi,Cornalba:2009ax,Hatta:2007cs,Bu:2011my}. In \cite{BallonBayona:2007qr} the soft wall AdS/QCD model was considered for scalar hadrons and a hybrid model involving a soft wall cut off for the photons and a hard wall cut off for the fermions was also discussed.

However, in neither of these articles the DIS was studied for baryons satisfying the experimentally observed mass spectrum for nucleons. So, the important case of the determination of the structure functions for baryons
with a mass spectrum consistent with the physical observations was still lacking. 

The purpose of this paper is to fill this gap by considering baryons in an AdS/QCD soft wall model with dressed mass in order to describe DIS in the AdS/QCD context but with baryons that present a spectrum similar to the physically observed one. We will use the modified soft wall model studied in \cite{Abidin:2009hr,Gutsche:2011vb} 
to describe the baryons and then calculate their DIS structure functions. 
Once we have the structure functions for these baryons we will compare our $F_2$ with experimental results for the proton in the regime of large values of the Bjorken parameter $x$ where the supergravity approximation that we use is more reliable.

The present paper is focused in large x limit, but it is important say that Bottom Up holographical models had been used in small x limit too in a successful way as you can see for example in \cite{Polchinski:2002jw, BallonBayona:2007rs, Cornalba:2010vk, Brower:2010wf}.

In section {\bf 2} we present a briefly review of  DIS and hadronic structure functions.
In section {\bf 3} we show our AdS/QCD calculation of the structure functions. Then in section {\bf 4} we
analyze our results, discussing the possible choices of the parameters of the model, plotting the structure function $F_2 $ for some kinematical regimes and comparing with experimental results.


\section{ Deep Inelastic Scattering and Structure Functions }

Deep inelastic scattering (DIS) is a process where a highly energetic lepton, an electron in general, 
interacts with a hadronic target through the exchange of a virtual photon, as show in the diagram of figure {\bf 1}.
The momenta of the photon and of the initial hadron are respectively $q^\mu$ and $P^\mu $. After the interaction  there is a final hadronic state represented by $X$ with momentum $P_X^\mu$. 
The experimental measurement of the inclusive cross section of DIS corresponds to detecting the final lepton, thus determining the momentum transfer $q^\mu$, but not the final hadronic state $X$. That means, summing over all possible final hadronic states $X$.     
One usually parametrizes DIS using as dynamical variables the photon virtuality $q^2$ and the Bjorken parameter

\begin{equation} 
x \equiv  -q^2 /2P\cdot q \,.
 \label{Bjorkenx}
\end{equation}

\noindent Deep inelastic scattering, in the strict sense, corresponds to 
the limit $q^2\to\infty$, with $x$ fixed. 

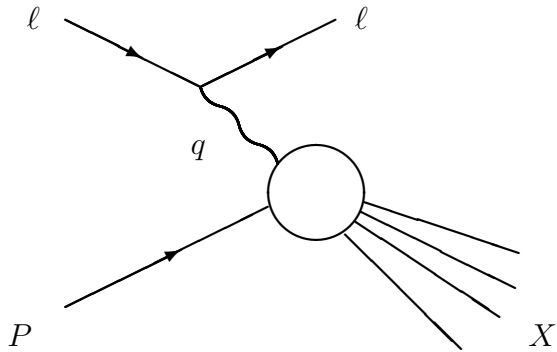
\begin{figure}\begin{center}
\setlength{\unitlength}{0.1in}
\vskip 3.cm
\begin{picture}(0,0)(15,0)
\rm
\thicklines
\put(1,14.5){$\ell$}
\put(3,15){\line(2,-1){7}}
\put(3,15){\vector(2,-1){4}}
\put(18,14.5){$\ell$}
\put(17,15){\line(-2,-1){7}}
\put(10,11.5){\vector(2,1){4.2}}
\put(9.5,8){$q$}
\bezier{300}(10,11.5)(10.2,10.7)(11,10.5)
\bezier{300}(11,10.5)(11.8,10.3)(12,9.5)
\bezier{300}(12,9.5)(12.2,8.7)(13,8.5)
\bezier{300}(13,8.5)(13.8,8.3)(14,7.5)
\put(0,-2){$P$}
\put(3,0){\line(2,1){10.5}}
\put(3,0){\vector(2,1){6}}
\put(16,6){\circle{5}}
\put(27,-2){$X$}
\put(18.5,5.5){\line(3,-1){8}}
\put(18.3,5){\line(2,-1){8}}
\put(18,4.5){\line(3,-2){7.5}}
\put(17.5,3.8){\line(1,-1){6}}
\end{picture}
\vskip 1.cm
\parbox{4.1 in}{\caption{Diagram of a deep inelastic scattering.}  }
\end{center}
\end{figure}
\vskip .5cm

The inclusive  cross section for DIS can be calculated from the hadronic tensor that 
is defined as 
\begin{equation}
W^{\mu\nu} \, = \frac{1}{8 \pi} \sum_s \, \int d^4y\, e^{iq\cdot y} \langle P, s \vert \, \Big[ J^\mu (y) , J^\nu (0) \Big] 
\, \vert P, s  \rangle \,,
 \label{HadronicTensor}
\end{equation}

\noindent where $ J^\mu(y)$ is the electromagnetic hadronic  current and $s$ the spin of the initial hadron. For an unpolarized scattering (spin independent) this tensor can be decomposed in terms of the  structure functions $F_1 (x,q^2) $ and $F_2 (x,q^2) $ as \cite{Manohar:1992tz}
\begin{equation}
W^{\mu\nu} \, = \, F_1 (x,q^2)  \Big( \eta^{\mu\nu} \,-\, \frac{q^\mu q^\nu}{q^2} \, \Big) 
\,+\,\frac{2x}{q^2} F_2 (x,q^2)  \Big( P^\mu \,+ \, \frac{q^\mu}{2x} \, \Big) 
\Big( P^\nu \,+ \, \frac{q^\nu}{2x} \, \Big)\, .\label{Structure}
\end{equation}

Considering only final states with just one baryon with mass $M_X$, we can introduce a basis of such  one particle states and write the hadronic tensor as    
 
\begin{eqnarray}
\label{hadronic2}
W^{\mu\nu} & = & \frac{1}{8 \pi} \sum_{s,s_X}  \sum_{M_X} \int \frac{d^4P_X}{2\pi^3)} \theta (P^0_X )
\delta \Big( P^2_X +  M_X^2 \, \Big) (2\pi )^4 \delta^4 (P +q - P_X ) \cr
&\times& \langle P,s \vert J^\nu ( 0 )   \vert P_X , s_X  \rangle\,
\langle P_X ,\, s_X  \vert J^\mu ( 0 )   \vert P,\, s \rangle\,
\cr
& =& \frac{1}{4} \sum_{s,s_X}  \sum_{M_X} \,\delta \Big( M_X^2 + (P+q)^2 \, \Big) 
\langle P,s \vert J^\nu ( 0 )   \vert P + q, s_X  \rangle\,
\langle P + q,\, s_X  \vert J^\mu ( 0 )   \vert P,\, s \rangle\,\,.\cr & &
\end{eqnarray}

So, in order to calculate the hadronic structure functions one needs to find the matrix elements of the current
and also the spectrum of hadronic masses $M_X\,$.
We will present in the next section the calculation of these quantities using the modified soft wall model.

\section{ DIS in the soft wall model with fermionic dressed bulk mass} 
In the soft wall model, a four dimensional gauge theory is represented by a gravity dual consisting
of fields living in anti-de Sitter space $AdS_5 $, whose metric can be written as 

\begin{equation}
\label{AdS} ds^2 \equiv g_{mn} \,dy^m dy^n \,= \, \frac{R^2}{z^2}( dz^2 +
\eta_{\mu\nu} dy^\mu dy^\nu  )\,\,= \, \frac{R^2}{z^2}( dz^2 - dt^2
+ (d\vec y )^2 )\,\,, 
\end{equation}

\noindent with the presence of an additional background of the form $e^{-\kappa^2 z^2}$. This factor plays the role of an infrared cut off, with the energy parameter $\kappa$ representing the corresponding scale. The general form of an action integral corresponding to a given Lagrangian 
density ${\cal L}$ is
 
\begin{equation}
I \,=\, \int dz d^{4}y \, \sqrt{-g} \,\, e^{-\kappa^2 z^2} \, {\cal L}\,\,.
\label{action}
\end{equation}
  
For the case of scalar and vector fields, this kind of action integral leads to  dual four dimensional theories 
with infrared cut off and discrete mass spectra. For fermions one must do some modification in order to find
a dual theory with an infrared cut off. This happens because using a standard fermionic Lagrangian in the action above the background $e^{-\kappa^2 z^2}$ factors out of the equations of motion. So, the fermions are not affected by the energy scale $\kappa$ and get no discrete mass spectrum. This problem can be solved introducing a $z$ dependent mass, as studied in \cite{Abidin:2009hr,Gutsche:2011vb}.
Following this approach, the appropriate action that describes the dynamics of the fermionic and gauge fields and their interaction is

\begin{eqnarray}
\label{Action1}
I &=& \int dz d^{4}y \, \sqrt{-g} \, e^{- \kappa^2 z^2 }  \,\biggl[ -
 \frac{1}{4} F_{mn} F^{mn} \, + \,  \frac{i}{2} \bar\Psi \epsilon_a^m \Gamma^a 
{\cal D}_m \Psi \cr &-&  \frac{i}{2} ({\cal D}_m\Psi)^\dagger \Gamma^0 
\epsilon_a^m \Gamma^a \Psi \,- \,  \bar\Psi \Big(\mu + V_F(z)\Big) 
\Psi \biggr] \,,
\end{eqnarray}
 
 \noindent where the field strength is $ F^{mn} = \partial^m A^n - \partial^n A^m \,$ and
$\epsilon_a^m = \delta_a^m \frac{z}{R} $ and the covariant derivative is: 
\begin{equation} 
{\cal D}_m = \partial_m - \frac{1}{8} \omega_m^{ab} \, [\Gamma_a, \Gamma_b] - i g_{_{V}} A_m\,,
\end{equation}
\noindent with: $ \omega_m^{ab} = - \frac{1}{z} (\delta^a_z \delta^b_m - \delta^b_z \delta^a_m)\,$.
The coupling $g_{_{V}}$ will be associated with the electric charge $e$  and  $\Gamma^a=(\gamma^\mu, - i\gamma^5)$ are the Dirac matrices. 

In order to produce a discrete spectrum with the appropriate spectrum for fermions we introduced the effective fermionic potential $V_F(z) = \kappa^2 z^2 /R$ depending on the $z$ coordinate 
(associated with the boundary energy scale), to dress the fermionic bulk mass.
Note that we are using here the same parameter $\kappa$ that appears in eq. (\ref{action}) in the dilaton  background.
We do this because in both cases the parameter $\kappa$ has the role of an infrared regulator that dictates the slope in a plot $m^{2}$ as a function of n (radial quantum number), as the hadronic mass spectrum suggest and it is widely accepted. The factor $\kappa$  in the potential will appear, as we will see, in spectrum of the fermions, while the factor $\kappa$ in the dilaton background appears in the gauge field solution and would 
also appear in the spectrum of vector mesons, as studied in \cite{Karch:2006pv}. So we take $\kappa$ as some universal infrared mass scale of the model.

The physical motivation for using a potential depending on the $z$ coordinate is the following. On one hand, according to the AdS/CFT dictionary, the mass of a supergravity bulk field is related  to the dimension of the corresponding  boundary operator.    
On the other hand, in general the dimensions of quantum operators  receive anomalous contributions, depending on the  energy scale. Since the energy scale of the boundary theory is holographically related to the localization in the $z$ coordinate, the possibility of anomalous contributions to the dimension of the operators can be translated into z dependent masses for the dual bulk modes \cite{Vega:2008te,Cherman:2008eh}. 
This kind of procedure makes it possible to introduce an important ingredient of QCD that is not considered in most of the AdS / QCD models. Other related works that consider masses varying in the bulk can be found in, for example,  \cite{Forkel:2008un,Forkel:2010gu,Vega:2010ne,Vega:2011tg}.

For the gauge fields it is convenient \cite{Polchinski:2002jw} to impose the gauge condition
\begin{equation}
\label{gaugechoice}
 \partial_\mu A^\mu \,+\,
z e^{ \kappa^2 z^2} \partial_z \Big( e^{-\kappa^2 z^2} \frac{1}{z}
 A_z \Big) \,=\,0 \,.
\end{equation}

Note that from now on we use the notation of raising and lowering the four dimensional indices with the 
Minkowski metric:  $A^\mu \equiv \eta^{\mu\nu} A_\nu\,$ and $\Box \equiv \eta^{\mu\nu}\partial_\mu \partial_\nu\,$. 
With the gauge choice (\ref{gaugechoice}) the equations of motion that emerge from the action (\ref{Action1}) are
\begin{eqnarray}
\label{Solutionsgauge}
\Box  A^\mu  &+& z e^{\kappa^2 z^2}  \partial_z \Big( e^{-\kappa^2 z^2} \frac{1}{z}
\partial_z A^\mu \Big) \,=\, 0\nonumber\\
\Box A_z &-&  \partial_z \Big( \partial_\mu A^\mu \Big) \,=\, 0\,.
\end{eqnarray}

\noindent We impose the condition that the boundary value of the gauge field represents a virtual photon with polarization $\eta^\mu$ and  space-like momentum $ q^\mu$ 
\begin{equation}
A_\mu (z, y) \vert_{z\to 0} \,=\, \eta_{\mu} \, e^{iq\cdot y} \,,
\end{equation}

\noindent  The corresponding solutions are 
\begin{eqnarray}
A_\mu (z, y) &=& \eta_\mu \, e^{iq\cdot y} \,\kappa^2 \,\,  \Gamma (1+\frac{q^2}{4\kappa^2} )\,\, z^2 \,\,{\cal U} (1+\frac{q^2}{4 \kappa^2} ; 2 ; \kappa^2 z^2 )
\nonumber\\
A_z (z, y)  &=& \frac{i}{2} \,  \eta \cdot q \,  e^{iq\cdot y} \,\, \Gamma (1+\frac{q^2}{4\kappa^2 } )\,\, z \,\, 
{\cal U} (1+\frac{q^2}{4 \kappa^2} ; 1 ; \kappa^2 z^2 )\,, 
\label{Gauge}
\end{eqnarray} 

\noindent where $\,{\cal U} (a;b;w) \,$ are the confluent hypergeometric functions of the second kind.  

Now, regarding the fermionic fields, it is convenient to do the following field 
redefinition

\begin{equation} 
\Psi (y,z) = e^{ + \kappa^2 z^2/2} \psi(y,z).
\end{equation} 

\noindent in such a way that the equations of motion take the form 
 
\begin{equation}
\biggl[ i\not\!\partial + \gamma^5\partial_z 
- \frac{2}{z} \gamma^5 
-  \frac{1}{z} \Big(m +  \kappa^2 z^2 \Big)  \biggr] \psi(y,z) = 0 \,, 
\end{equation} 

\noindent where $ \not\!\partial = \gamma^\mu \, \partial_\mu$. 
The quantity $\mu$, the bulk fermion mass was replaced by the dimensionless parameter $m = \mu R $. 
In order to solve the equation of motion, we decompose the fermionic field into chiral components 

\begin{equation}
\psi(y,z) = \psi_L(y,z) + \psi_R(y,z)\,, \quad 
\psi_{L/R} = \frac{1 \mp \gamma^5}{2} \psi \,. 
\end{equation} 
\noindent with $(\gamma^5)^2 = 1 $.

We will consider solutions of the form

\begin{equation}
\psi_{L/R}(y,z) = e^{iP\cdot y } \frac12 \Big( 1 \mp \gamma^5 \Big) u_s(P) \, \frac{z^2}{R^2}
f_{L/R}(z) \,, 
\end{equation} 

\noindent that contain the plane wave factor and the four component spinor $ u_s(P) $ corresponding to a four dimensional free fermion with momentum $P^\mu $ and spin $s$.  Then, the $z$ dependent parts of chiral components of
the fields: $f_{L/R}(z)$ must satisfy  

\begin{equation}
\label{EqzLR}
\biggl[ -\partial_z^2 
+ \kappa^4 z^2 + 2 \kappa^2 \Big(m \mp \frac{1}{2} \Big) 
+ \frac{m (m \pm 1)}{z^2} \biggr] f_{L/R}(z) = - P^2  \, f_{L/R}(z) \,.
\end{equation}

Following the usual prescription of gauge/gravity dualities, normalizable solutions for fields in the gravity side are dual to states in the boundary four dimensional theory. Equations (\ref{EqzLR}) have normalizable solutions only when $- P^2$, the four dimensional mass,  has the discrete values 
\begin{equation} 
\label{Masses}
- P^2_n = M_n^2 = 4 \kappa^2 \Big( n + m + \frac{1}{2} \Big) \,, 
\end{equation}  

\noindent with $ n= 0,1,2,..., $. The corresponding discrete set of normalizable solutions  $f^n_{L/R}(z)$ are 
\begin{eqnarray} 
f^n_{L}(z) &=& \sqrt{\frac{2\Gamma(n+1)}{\Gamma(n+m+3/2)}} \ \kappa^{m+3/2}
\ z^{m+1} \ e^{-\kappa^2 z^2/2} \ L_n^{m+1/2}(\kappa^2z^2) \,, \\
f^n_{R}(z) &=& \sqrt{\frac{2\Gamma(n+1)}{\Gamma(n+m+1/2)}} \ \kappa^{m+1/2}
\ z^{m} \ e^{-\kappa^2 z^2/2} \ L_n^{m-1/2}(\kappa^2z^2)
\end{eqnarray}
 
\noindent with normalization condition $\int\limits_0^\infty dz \, f^{n^\prime}_{L/R}(z) f^n_{L/R}(z) = \delta_{n^{\prime}n}\,$.  

We want to describe processes where the initial state is a proton that absorbs a virtual photon transforming
into a final state corresponding to an excited hadronic state of spin $1/2$. So we consider
the interaction action

\begin{eqnarray}
\label{InteractionAction}
S_{int} [i,X] =  g_{_{V}}\, \int dz d^{4}y \sqrt{-g} e^{- \kappa^2 z^2} \,\frac{z}{R}
\, A_m \bar\Psi_X \, \gamma^m \, \Psi_i \,.
\end{eqnarray}

\noindent where $\Psi_i$ represents the initial proton, that we take as the state with lowest mass level, corresponding to $ n=0$.
So the initial momentum $P_i \equiv p $ satisfies $-p^2 = M_0^2 = 4 \kappa^2 ( m + 1/2 )$. 
The fermionic field $\Psi_X$ represents a final state with (higher) mass $M_X$ and momentum $P_X = p + q $  satisfying $ - P_X^2 = M_X^2  = 4 \kappa^2 ( n_X + m + 1/2 ) $, where we are representing as $n_X$ the integer
associated with the excitation level of the final state. The corresponding solutions have the form 

\begin{eqnarray}
\label{SolutionFermions}
\Psi_i &=& e^{ip\cdot y } e^{ + \kappa^2 z^2/2} \, \frac{z^2}{R^2} \Big[ \Big(\frac{1 - \gamma^5}{2} \Big) 
u_{s_{i}}(p) \, 
f^0_{L}(z) \,+\, \Big( \frac{1 + \gamma^5}{2} \Big) u_{s_{i}} (p) \,f^0_{R}(z) \Big] \cr \cr
\Psi_X &=& e^{iP_X \cdot y } e^{ + \kappa^2 z^2/2} \, \frac{z^2}{R^2} \Big[ \Big(\frac{1 - \gamma^5}{2} \Big) 
u_{s_{X}}(P_X) \, f^{n_X}_{L}(z) \,+\, \Big( \frac{1 + \gamma^5}{2} \Big) u_{s_{X}} (P_X) \,f^{n_X}_{R}(z) \Big]\,\,,
\cr
& &
\end{eqnarray}

\noindent where $s_i $ and $s_X$ are the spins of the initial and final fermionic states. 

In order to find out the structure functions for the hadrons we have to calculate
the interaction action (\ref{InteractionAction}) with the solutions for the fields. 
We can simplify the calculations by considering, as it was done in \cite{Polchinski:2002jw,BallonBayona:2007qr},
that we are probing the hadron with a particular photon with polarization $\eta^\mu $ satisfying $\eta \cdot q = 0$. 
For this situation the $z$ component of the gauge field does not contribute and, substituting the solutions 
(\ref{Solutionsgauge}) and (\ref{SolutionFermions}) in (\ref{InteractionAction}) the interaction action, {\underline \it for this initial and final states $(i,X)$ } takes the 
on shell form
 
\begin{eqnarray}
\label{InteractionAction2}
S_{int} [i,X] &=&  \frac{g_{_{V}}}{2}\, \, (2\pi )^4 \delta^4 ( P_X -p - q ) \eta_\mu \Big[ \bar{u}_{s_{X}}(P_X) \gamma^\mu \Big(\frac{1 - \gamma^5}{2} \Big) u_{s_{i}}(p) {\cal I}_L (n_x ) \cr & & +  \,\bar{u}_{s_{X}}(P_X) \gamma^\mu \Big(\frac{1 + \gamma^5}{2} \Big) u_{s_{i}}(p) {\cal I}_R (n_x ) \Big]   \,,
\end{eqnarray}

\noindent where $ {\cal I}_L (n_x ) $ and ${\cal I}_R (n_x ) $ are integrals involving the two fermionic solutions with different chiralities. They have a similar structure and can be written, in terms of the variable $ w \equiv \kappa^2 z^2 $, in the general form:

\begin{equation} 
\label{generalintegral}
{\cal I} ( {\bar m}, n_X )  =  C (  {\bar m}, n_X  ) \, 
\Gamma( 1 + \frac{q^2}{4 \kappa^2} ) \, \int_0^\infty dw w^{{\bar m} -1} e^{-w} {\cal U} (1+\frac{q^2}{4\kappa^2} ; 2 ; w ) \,  L_{n_X}^{{\bar m} - 2}( w )\,,
\end{equation} 

\noindent where 
\begin{equation} 
C ( {\bar m}, n_X )  =  \sqrt{\frac{4 \Gamma(n_X +1)}{\Gamma({\bar m} -1 )\Gamma(n_X + {\bar m} -1 )}}\,. 
\end{equation} 

The integrals  $ {\cal I}_L (n_x ) $ and ${\cal I}_R (n_x ) $ correspond  to 
${\cal I} ( {\bar m}, n_X )$ with $ {\bar m} = m + 5/2 $ and $ {\bar m} =  m + 3/2 $ respectively.  
Performing the integral (\ref{generalintegral}) we get
\begin{equation}
\label{I}
{\cal I} (\bar{m},n_{X})=\frac{q^2 \Gamma \left(\bar{m}\right) \sqrt{\frac{\Gamma \left(\bar{m}-1\right) \Gamma \left(n_X +\bar{m}-1\right)}{\Gamma (n_X +1)}} \Gamma \left(\frac{q^2}{4 \kappa ^2}+n_X\right)}{2 \kappa ^2 \Gamma \left(\bar{m}-1\right) \Gamma \left(\frac{q^2}{4 \kappa ^2}+n_X+\bar{m}\right)}
\end{equation}

Now that we calculated the action in the soft wall background for photons interacting with fermions with dressed mass, we connect this result with the four dimensional boundary theory using a proposal similar to the one used in refs. \cite{Polchinski:2002jw,BallonBayona:2007qr}.
In these references the matrix element of the fermionic electromagnetic current on the boundary four dimensional theory was considered to be equal to
the corresponding bulk interaction action. Here we will take a different point of view and consider that bulk/boundary duality implies that these quantities are proportional, rather than necessarily equal. So we assume the relations:

\begin{eqnarray}
\eta_\mu \langle P_X  \vert {\tilde J}^\mu ( q )   \vert P  \rangle\,
&=&  (2 \pi)^4 \, \delta^4 ( P_X - P - q ) \,\eta_\mu \,  \langle P + q \vert J^\mu ( 0 )  \vert P \rangle
\,=\,{\cal K}_{_{eff}} \, S_{int} [i,X]\, \cr \cr
\eta_\mu \langle P  \vert {\tilde J}^\mu ( q )   \vert P_X  \rangle\,
&=&  (2 \pi)^4 \, \delta^4 ( P_X - P - q ) \,\eta_\mu \,  \langle P \vert J^\mu ( 0 )  \vert P + q \rangle
\,=\,\,{\cal K}_{_{eff}} \, S_{int} [X,i]\,,\cr & & 
\label{Prescription}
\end{eqnarray}

\noindent where ${\cal K}_{_{eff}}$ plays the role of a bulk/boundary effective factor that phenomenologically adjust the bulk supergravity 
quantities to the boundary observed ones.  
Following this prescription, the hadronic tensor in eq. (\ref{hadronic2}), contracted with the photon polarization $\eta $,  can then be written in terms of our interaction action of 
eq. (\ref{InteractionAction2}) as 

\begin{eqnarray}
\label{Amplitudewithspinors}
& &  \eta_\mu \eta_\nu W^{\mu\nu} \, = \,\frac14 \,\sum_{M_X} \delta ( M_X^2 + (p+q)^2 ) \,\frac{g_{eff}^2}{4} \times
\cr & &   \sum_{ s_i } \sum_{s_X} \, \Big\{ {\cal I}^2_L {\cal I}^2_R  
 \Big( \bar{u}_{s_{X}} \gamma^\mu \Gamma^{(- )} u_{s_{i}} \bar{u}_{s_{i}} \gamma^\nu \Gamma^{(+ )} u_{s_{X}} + \bar{u}_{s_{X}} \gamma^\mu  \Gamma^{(+)} u_{s_{i}} \bar{u}_{s_{i}} \gamma^\nu \Gamma^{(- )} u_{s_{X}} \Big)
\cr \cr 
& & +  {\cal I}^2_L  \bar{u}_{s_{X}} \gamma^\mu \Gamma^{(-)} u_{s_{i}} \bar{u}_{s_{i}} \gamma^\nu \Gamma^{(-)} u_{s_{X}}
\, + {\cal I}^2_R \bar{u}_{s_{X}}\gamma^\mu \Gamma^{(+)} u_{s_{i}} \bar{u}_{s_{i}} \gamma^\nu 
\Gamma^{(+)} u_{s_{X}}\Big\}
\end{eqnarray}

\noindent where we defined $ g_{eff} \equiv  {\cal K}_{_{eff}} \, g_{_{V}} $ and 
$ \Gamma^{(\pm )} \equiv \frac{1 \pm \gamma^5}{2}$ and omitted the dependence of the spinors on the momenta. Note that we are summing over the final spins and averaging over the initial ones.
Using the property

$$ \sum_s (u_s)_{_\alpha} (p) (\bar{u}_s)_{_\beta} (p) = (\gamma^\mu p_\mu + M )_{_{\alpha\beta}} \,,$$

\noindent  satisfied by both the initial and final state (on shell) spinors, with the corresponding masses and momenta, we find

\begin{eqnarray}
\label{Amplitude}
\eta_\mu \eta_\nu W^{\mu\nu}  &=&  \frac14 \, \sum_{M_X} \delta ( M_X^2 + (p+q)^2 ) \,  \, g_{eff}^2
\Big\{ - {\cal I}^2_L (n_x ) {\cal I}^2_R (n_x ) M_X  M_0 \,\eta \cdot \eta  \cr \cr 
& & + \Big(  {\cal I}^2_L (n_x ) + {\cal I}^2_R (n_x ) \Big) \Big(   ( p\cdot \eta )^2 - \frac{1}{2} ( p^2 + p \cdot q )\, 
\eta \cdot \eta \Big) \Big\}
\end{eqnarray}

The sum over the final states $X$ can be approximated by an integral over a continuum of states as it was done in
in refs. \cite{Polchinski:2002jw,BallonBayona:2007qr}. In the present case we this corresponds to replacing the
sum over $X$ of the delta functions by the factor: $ \frac{1}{4 \kappa^2} $. That means:

\begin{eqnarray}
\label{ApproximateAmplitude}
\eta_\mu \eta_\nu W^{\mu\nu}  &\approx &  \frac{ g_{eff}^2}{16 \kappa^2}
\Big\{ - {\cal I}^2_L (n_x ) {\cal I}^2_R (n_x ) M_X  M_0 \,\eta \cdot \eta  \cr \cr 
& & + \Big(  {\cal I}^2_L (n_x ) + {\cal I}^2_R (n_x ) \Big) \Big(   ( p\cdot \eta )^2 - \frac{1}{2} ( p^2 + p \cdot q )\, 
\eta \cdot \eta \Big) \Big\}
\end{eqnarray}

For the particular photon that we are considering, with $\eta \cdot q = 0$ we get from eq. (\ref{Structure})

\begin{equation} 
 \eta_\mu \eta_\nu W^{\mu\nu}  \,=\, \eta^2  F_1  \,+\, \frac{2 x}{q^2 } 
  ( \eta \cdot p )^2 \, F_2 
\end{equation} 

Comparing this with our expression for the hadronic tensor eq. (\ref{ApproximateAmplitude}) we find our results for the fermionic structure functions

\begin{eqnarray}
\label{Result1}
 F_1  &=& \frac{ g_{eff}^2 }{16  \kappa^2  } 
\Big\{ \Big(  {\cal I}^2_L (n_x ) + {\cal I}^2_R (n_x ) \Big) \Big( \frac{M_0^2}{2 } + 
\frac{q^2}{4 x} \Big) \cr \cr & & - {\cal I}_L (n_x ) {\cal I}_R (n_x ) M_0
 \sqrt{ M_0^2 + \frac{q^2(1-x)}{x} } \Big\}\cr \cr
 F_2 & = & \frac { g_{eff}^2 \,}{32 \kappa^2 } 
 \Big(  {\cal I}^2_L (n_x ) + {\cal I}^2_R (n_x ) \Big) \frac{ q^2 }{  x}  
\end{eqnarray}

The excitation level $n_x$ of the final state is not an independent variable. It can be expressed in terms 
the DIS variables $q^2$ and $x$. Using eqs.  (\ref{Bjorkenx}) and (\ref{Masses}) one finds

\begin{equation} 
\label{nx1}
- (p + q)^2 = M_0^2 + q^2 \Big[ \frac{1}{x} - 1 \Big]\,= M_X^2 \,=   4 \kappa^2 \Big( n_X + m + \frac{1}{2} \Big) 
\,, 
\end{equation}  
\noindent so
\begin{equation} 
\label{nx2}
n_X \,=\,  \frac{q^2}{4 \kappa^2 } \Big[ \frac{1}{x} - 1 \Big]
\,, 
\end{equation}  

\begin{center}
\begin{figure}[ht]
~~~~~~~~~~~~~~~~~~~~~~~~~~~~\includegraphics[width=2.8 in]{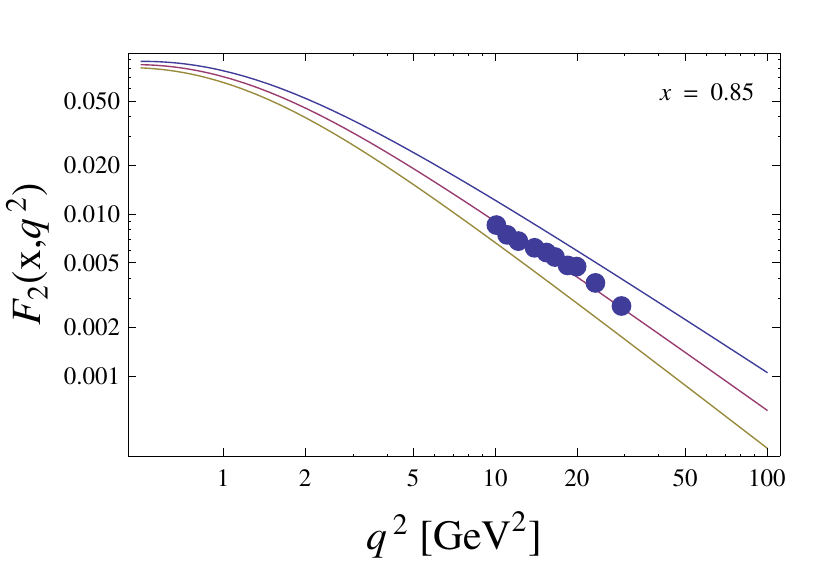}
\caption{$F_{2}$ as a function of $ q^2$, where $q^2$ is in units of (GeV)$^2$. The plot consider $x = 0.85$. The higher curve in for $m = 0.6$, the middle one for $m = 0.7$ and the lower is plotted with $m = 0.8$. Dots correspond to experimental data \cite{Nakamura:2010zzi, Whitlow:1991uw}.}
\end{figure} 
\end{center}

\section{ Analysis of the results }

In order to calculate the structure functions and analyze their dependence on $x$ and $q^2 $, first we have to fix the parameters of the model. The effective coupling $ g_{eff} \equiv  {\cal K}_{_{eff}} \,g_{V}$ contains the electric charge $g_{_{V}}$, that satisfies $ g_{_{V}}^2 = 1/137 $ but multiplied by the yet undetermined parameter $ {\cal K}_{_{eff}} $ inserted in the model in order to  phenomenologically adjust the relation between bulk and boundary quantities and essentially fix the size the structure functions. 
We found that the choice $g_{eff}^{2} = 0.66$ leads to structure functions $F_{2}$ with values compatible with experimental data. So we used this value in the calculations.  

The infrared energy scale $\kappa$ is related to the slope in a plot of $m^{2}$ as a function of $n$. In fact, in the present holographic model these slope is $4 \kappa^{2}$. We will choose $4 \kappa^{2} = 0.9 \, GeV^{2}$ that was found in ref.
 \cite{Forkel:2007cm,Vega:2008te} to give a good adjustment for the nucleon masses. So we use $\kappa = 0.474 GeV$.
 
The other parameter of the model: $m$ is considered as a free parameter, associated, as we discussed before, with the anomalous dimension of the operator that creates the baryonic states. This parameter appears in mass the spectrum and can be fixed in this way. For example considering the  Proton mass equal to 0.938 GeV, m must be 0.477, but we prefer to consider values such that we get the best fit of the shape of the structure function $F_2 $ to the experimental data, and that produce values for the proton mass close to the experimental one.
The m values considered are 0.6, 0.7 and 0.8 that produce 0.995, 1.039 and 1.081 GeV respectively for the proton mass, and they were adjusted using data for x=0.85, see Fig. 2.
  
It is important to remark that the supergravity approximation used in the model is, in principle, valid only for large values of $x$. This happens because, as discussed in \cite{Polchinski:2002jw}, for low values of $x$ string theory corrections would become relevant. So we analyzed the region $ 0.8 < x < 1.0 $.
We show in Fig. 2 the structure functions found with the holographic model for $x = 0.85$ compared with the corresponding experimental data that appear in PDG  \cite{Nakamura:2010zzi, Whitlow:1991uw}  considering the values of $m$ that led to the best fits, that means, $m$ close to 0.7.

\begin{center}
\begin{figure}[ht]
  \begin{tabular}{c c}
    \includegraphics[width=2.8 in]{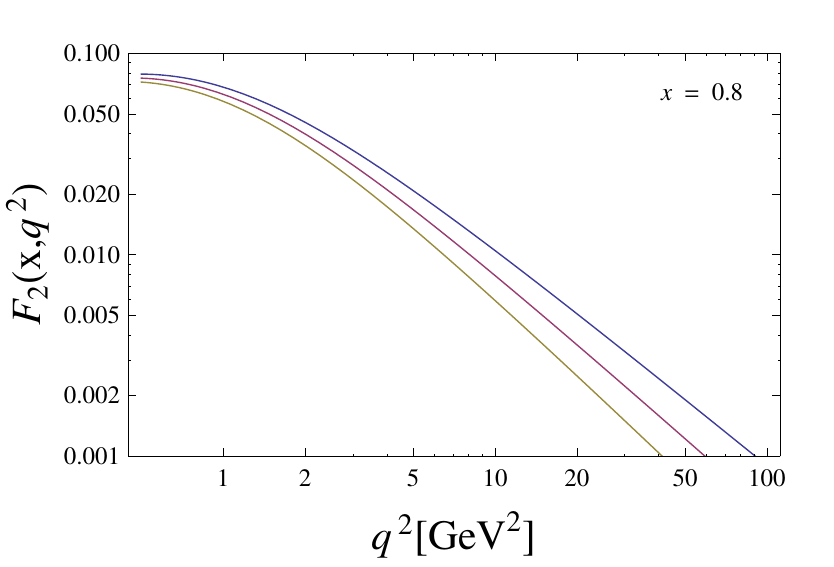} & \includegraphics[width=2.8 in]{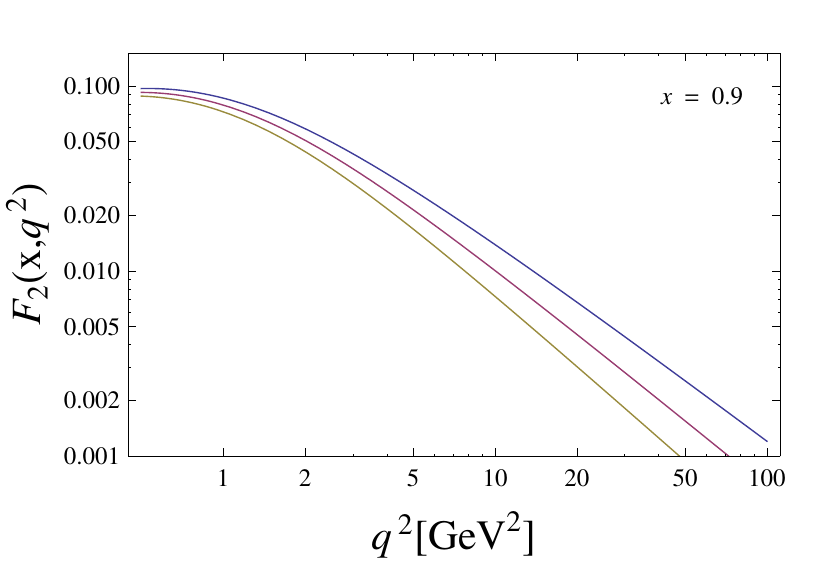}
  \end{tabular}
\caption{$F_{2}$ as a function of $ q^2$. The plot on the left consider $x = 0.8$ and on the right  $x = 0.9$. In both plots the higher curve in for $m = 0.6$, the middle one for $m = 0.7$ and the lower is plotted with $m = 0.8$.}
\end{figure} 
\end{center}

In order to check if the dependence of $F_{2}$ on $ q^2 $ has a similar form for other values of $x$ in the
range considered, we show in Fig. 3 this structure function for $x =0.8$ and $ x= 0.9$. In both cases we included plots for three different choices of the parameter m. We see that $F_{2}$ decreases with $q^2$ in a way that is compatible with the experimental results shown in figure (16.7) of \cite{Nakamura:2010zzi} or in \cite{Whitlow:1991uw} for $x=0.85$.   

It is important to stress the fact that we are only considering a region of high values of the Bjorken 
parameter $x$ (close to the elastic case $ x = 1$). For the range considered: $ 0.8 \le  x < 1 $ 
the experimental results show a strong dependence of the structure function $F_2$ on $q^2$. 
That means: there is no Bjorken scaling in this region. So, the picture of the virtual photon 
interacting with just one parton carrying a fraction $x$ of the hadron momentum, which corresponds 
to structure functions depending only on $x$ and not on $q^2$, does not hold in this region. 
It would be valid for smaller values of $x$, as can be seen in ref.  \cite{Nakamura:2010zzi}.   
This is consistent with our approximation of representing the final states by just a single hadron with 
the total momentum.

Now, let's consider the dependence of $F_{2}$ on the Bjorken parameter  $x$. There are not many experimental results for this structure function for $x$ close to one. But one finds in ref. \cite{Malace:2009kw} 
an interesting investigation of $F_{2}$ at large x, for low values of $q^2$, using 
some  parametrizations obtained from experimental data. In particular, they show results for the $F_2$ as a function of $x$ for the cases of $q^2 = 4 $ GeV$^2$   and  9 GeV$^2$. 
Their results show a decrease in this structure function when we increase $x$  in the interval $ 0.9 < x < 1$.  

We show in Fig. 4 the structure function $F_2 $ obtained using our holographic model with z dependent mass in the range $0.8 < x < 1.0 $, for these two values of $q^{2}$ analyzed in \cite{Malace:2009kw}, using values for m around 0.7. The order of magnitude of our results was adjusted by the choice of the effective coupling $g_{eff}$  to be consistent with the experimental values, so they are of the same order of those found in \cite{Malace:2009kw}, however, in contrast to the results of \cite{Malace:2009kw}, we found an increase in the structure functions when $ x \to 1 $. 
So, our model for DIS of baryons does not give a good description for the dependence of $F_2$  on $x$ at low 
$q^{2}$. This may be a consequence of the fact that in this simple model the final hadronic states have just one baryon with spin $1/2$. In a general non elastic process the final state may include more than one hadron and also
hadrons with higher spins. A description of such kind of process with multiple hadrons and also higher spins
is out of the scope of the present kind of model. Nevertheless it is interesting to see that the model is able to  reproduce, by adjusting the free parameters, the experimental dependence of $F_2 $ on $q^2 $. 

It is also interesting to consider the high q limit of the structure functions eq.(\ref{Result1}). Using the property  

\begin{center}
\begin{figure}[ht]
  \begin{tabular}{c c}
    \includegraphics[width=2.8 in]{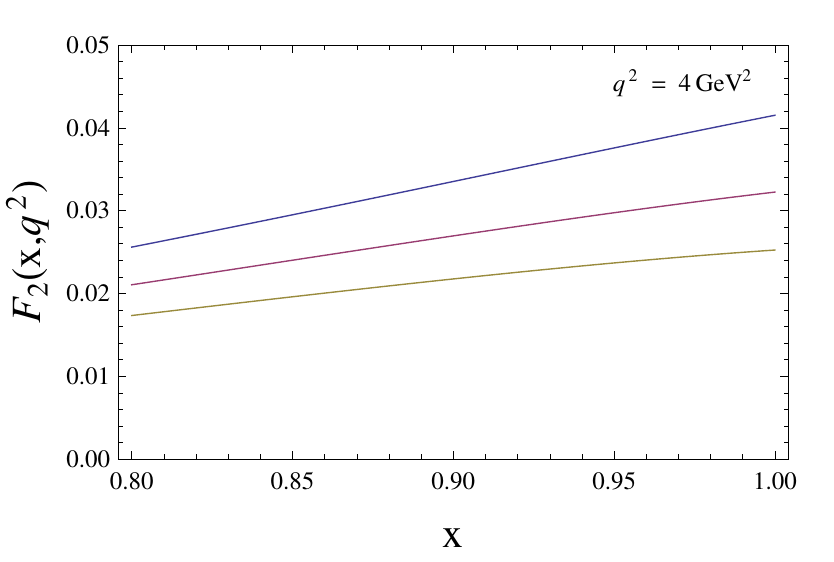} & \includegraphics[width=2.8 in]{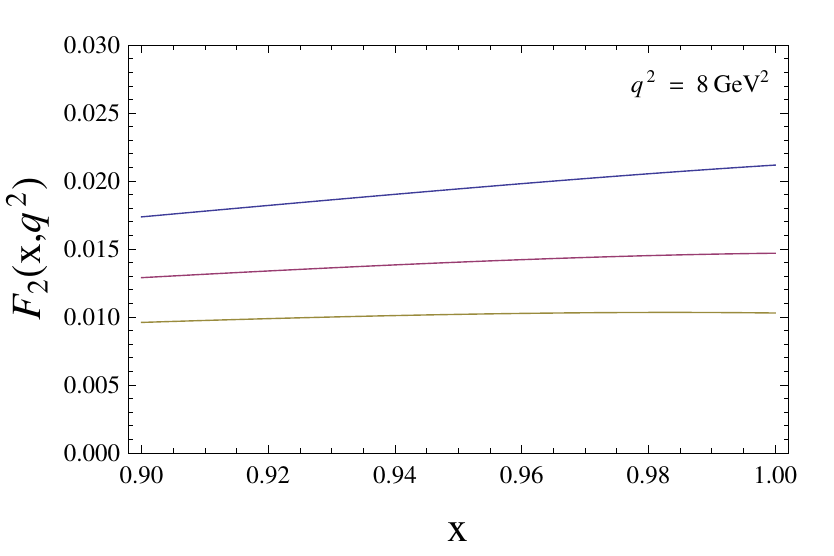}
  \end{tabular}
\caption{$F_{2}~v/s~x$. The plot at the left consider $q^{2} = 4 GeV^{2}$ and the other is for $q^{2} = 8 GeV^{2}$. In both plots the higher curve in for $m = 0.6$, the middle one for $m = 0.7$ and the lower is plotted with $m = 0.8$.}
\end{figure} 
\end{center}

\begin{equation} 
\label{Aprox}
\frac{\Gamma (a + y)}{\Gamma (b + y)} = \biggl( \frac{1}{y} \biggr)^{b-a} \biggl( 1 + O (y^{-1}) \biggr);~~~~~y\rightarrow \infty \,,
\end{equation}  

\noindent we can expand the integrals $I_{L,R}$ defined by eq. (\ref{I}) and that appear in eq.(\ref{Result1}). This way we find properties shared with other AdS/QCD models at dominant order in q, like 

\begin{equation} 
F_{2} = 2 F_{1}
\end{equation}
\noindent that implies that for $ x \to 1 $ we approach the Callan-Gross relation. We also find

\begin{equation} 
F_{2} \sim \biggl( \frac{q^{2}}{4 \kappa^{2}} \biggr)^{-m  - \frac{1}{2}} x^{m + \frac{5}{2}} (1 - x)^{m - \frac{1}{2}}.
\end{equation} 

\begin{center}
\begin{figure}[ht]
  \begin{tabular}{c c}
    \includegraphics[width=2.8 in]{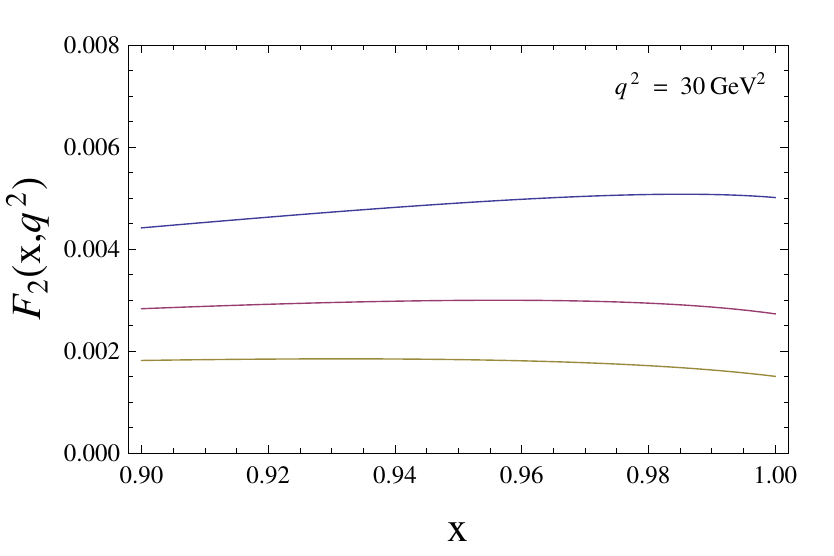} & \includegraphics[width=2.8 in]{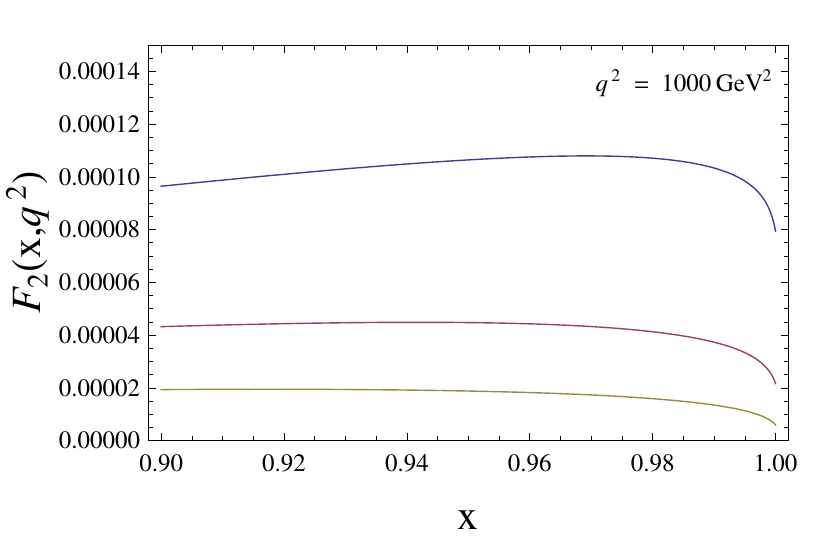}
  \end{tabular}
\caption{$F_{2}~v/s~x$. The plot at the left consider $q^{2} = 30~GeV^{2}$ and the other is for $q^{2} = 1000~GeV^{2}$. In both plots the higher curve in for $m = 0.6$, the middle one for $m = 0.7$ and the lower is plotted with $m = 0.8$. It can be seen that at high q the structure functions decrease for $x \to 1$.}
\end{figure} 
\end{center}

Notice that when $m = \tau - 3/2$, where $\tau$ is the twist (dimension of the operator that creates the state, minus the spin of the state),  our model reproduces results found in previous works that consider DIS in holographic models \cite{Polchinski:2002jw, BallonBayona:2007qr}, this is not strange, because the dilaton decouples in this limit, a property discused in \cite{Brodsky:2007hb, Vega:2012iz}.
Additionally, at high values for $q^{2}$, we find a decreasing behavior for the structure function when x is close to one, as can be seen in Fig. 5.

Finally, it interesting to discuss the role of the parameter $m $ ($ \equiv \mu R)$  that was used to adjust the form of the structure functions and find a nice fit to the experimental data. This parameter was introduced as a five dimensional mass term of the fermionic field (in units of the inverse of the AdS radius). 

The five dimensional mass in the gauge/string correspondence is related to the scaling dimension of the dual 
boundary operator. In the standard approach, based on AdS/CFT, a fermionic field with a five dimensional constant mass $\mu $ is dual to a boundary operator with a (constant) scaling dimension \cite{Polchinski:2002jw}

\begin{equation}
\label{Dimension}
\Delta = \mu R + 2 \,.
\end{equation}

\noindent 
Note that in a conformal field theory, that is the case in AdS/CFT, the dimensions of the operators do 
not vary with the energy.

Here, we followed a phenomenological AdS/QCD approach and described the fermionic fields as in refs.  \cite{Vega:2008te,Cherman:2008eh}.
That means, considering that in a non conformal theory the scaling dimensions of the operators in general vary with the energy scale, we introduced an effective mass for the fermionic field of the form: $ ( m + \kappa^2 z^2 )/R $. 
Since the bulk coordinate $z$ is related to the energy scale of the boundary theory, this $z$ dependent mass term represents an effective way of incorporating the anomalous dimension of the boundary fermionic operators into the model. 

The asymptotic behaviour of our fermionic solutions when $z \to 0$, that determines the dimension of the dual boundary operator,  is not affected by the presence of the  $ \kappa^2 z^2 $ term.
So, we can assume relation (\ref{Dimension}) to hold in our model. Thus, changing the value of $m$ corresponds to
changing the dimension of the baryon operator. 
We can illustrate this relation taking two limiting cases. First, if we consider a baryon as been build from 
three fermionic operators (the valence quarks) we should have $\Delta = 9/2$ and then $m = 5/2$.
On the other hand, if we consider the baryon operator as just one fermionic operator (like a particle without any internal structure) we have $\Delta = 1/2$ leading to $ m = -1/2$.
Note that these two values for $\Delta $ are just classical scaling dimensions.  
The values of $m$  that we found to give a nice fit for the structure function $F_2$ are in a region 
near $ m \approx 0.7$. We may interpret this result as indicating that the hadron behaves,
in the range of $x$ and $q^2 $ that we considered as having an effective scaling dimension $ \Delta_{effective} \approx 2.7 $.

\noindent {\bf Acknowledgments:} We would like to thank Carlos Alfonso Ballon Bayona for important discussions.
N.B. is partially supported by CNPq and Capes (Brazil)  and A.V is supported by Fondecyt (Chile) under Grant No. 3100028. A.V. is grateful for the hospitality of the Instituto de F\'{\i}sica of Universidade Federal do Rio de Janeiro, where this work started.

 \end{document}